\documentclass[10pt,prl,twocolumn]{revtex4}
\usepackage{graphicx}
\usepackage{amssymb}
\usepackage{times}
\usepackage{amsmath}

\begin{document}
\title{Asynchronous entanglement from coherently coupled nonlinear cavities.}

\author{Terry G. McRae$^{1,2}$, and Warwick P. Bowen$^1$}
\address{$^1$ School of Mathematics and Physics, University of Queensland, Brisbane, QLD 4072, Australia}
\address{$^2$ MacDiarmid Institute, Physics Department, University of Otago, Dunedin, New Zealand}

\begin{abstract}
The output fields of a pair of coherently coupled nonlinear optical cavities are found to exhibit strong optical entanglement. For sufficiently strong coupling the quantum correlations become asynchronous providing a resource for quantum information protocols such as all-optical quantum memories. A straightforward experimental implementation applicable to whispering gallery mode resonators such as microtoroids is proposed.
\end{abstract}


\date{\today} \maketitle

Last century, interest in entanglement was motivated predominately by its fundamental significance as ``{\it the} characteristic trait of quantum mechanics"\cite{Schrodinger}.
In recent years, however, a paradigm shift has occurred due to the realization that entanglement can facilitate powerful measurement, computation, and communication tasks\cite{Chuang}. Optical entanglement is of particular significance, both enabling applications such as quantum cryptography, quantum information networks, and quantum metrology\cite{Chuang}; and facilitating the most stringent tests of quantum mechanics performed to date\cite{Aspect}.

Here we investigate, for the first time, optical entanglement generated from a pair of coherently coupled nonlinear optical cavities. Coherently coupled optical cavities are of fundamental interest, enabling for example an all-optical analog to electromagnetically induced transparency (EIT)\cite{YanikPRL04A} with the capacity to both slow and stop light\cite{TotsukaPRL07}. We show that the introduction of
nonlinearity enables arbitrarily strong entanglement to be generated. Interestingly, however, for strong coupling, the entanglement is quenched, and a new form of asynchronous entanglement exhibiting time-delayed quantum correlations becomes apparent. This entanglement is analogous to that generated in $\lambda$-type atomic ensemble quantum memories\cite{Kuzmich03}, which have already proved to be an enabling technology in quantum information applications such as quantum repeaters and on-demand single photon sources\cite{DLCZ}.

In contrast to atomic ensembles, quantum memory capabilities are not typically available in nonlinear optical systems. Indeed, they are only present in our system due to the EIT-like nature of the coupled cavities; and to our knowledge have not been predicted for any other nonlinear optics based entanglement source. The asynchronous entanglement predicted here has many applications in quantum information science. For example, an all-optical quantum memory could be implemented utilizing asynchronous entanglement as the non-classical resource for quantum teleportation\cite{furusawa}.

\begin{figure}[t!]
\begin{center}
\includegraphics[width=8cm]{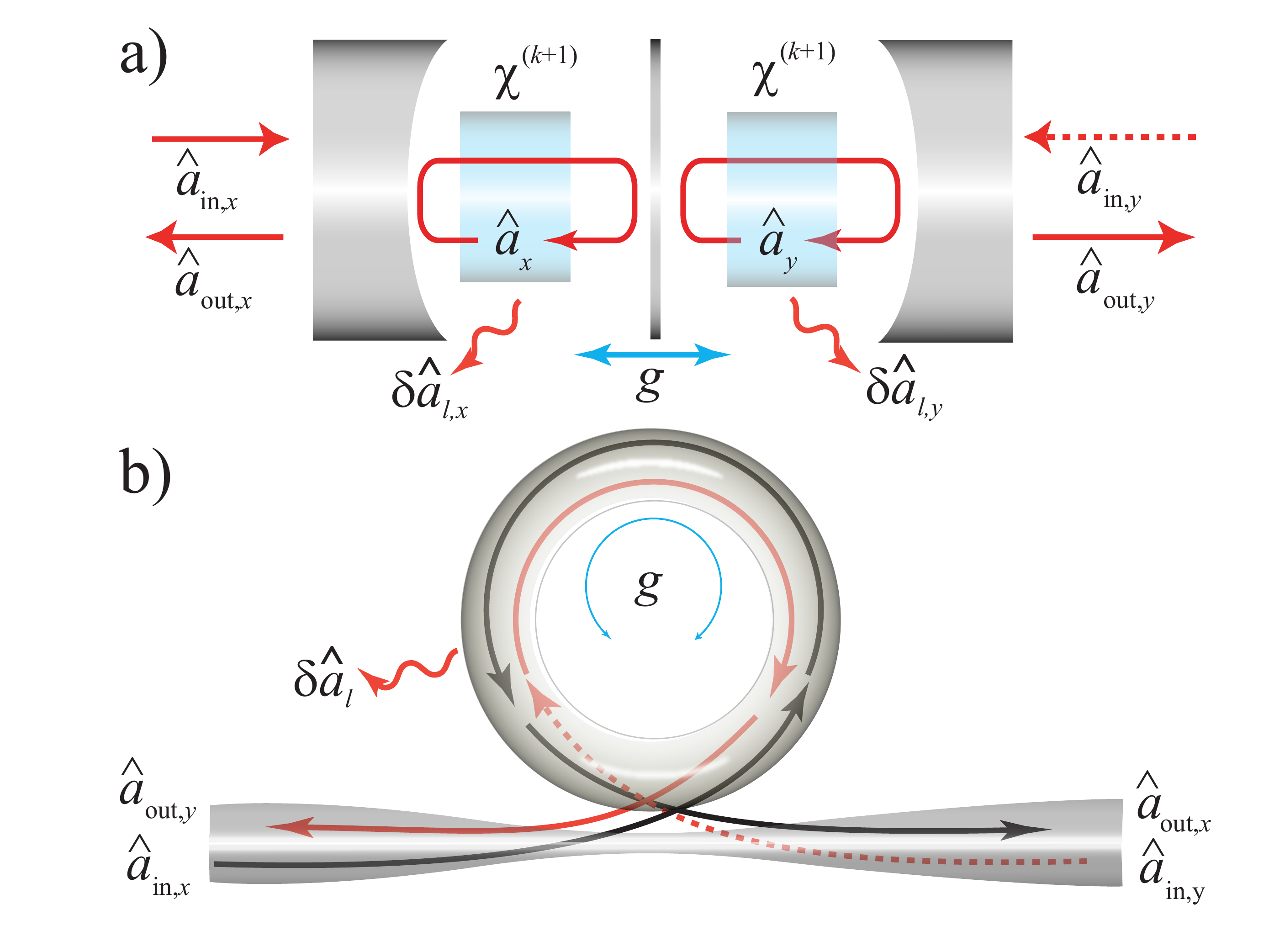}
\caption{ (color online). Schematics of coupled cavity systems. a) two optical cavities coupled via a partially reflective mirror, each containing a $(k \! + \! 1)^{\rm th}$ order nonlinear medium. b) an equivalent WGM resonator with two counter propagating optical modes coupled by optical scattering and intrinsic nonlinearity.}
\vspace{-8mm}
\label{fig1}
\end{center}
\end{figure}

 This Letter models a pair of identical below threshold coherently coupled optical parametric oscillators (OPOs) each consisting of a $\chi^{(k+1)}$ general nonlinear medium enclosed in an optical cavity as shown in Fig.~\ref{fig1}(a). Such a configuration is experimentally relevant, most notably to ultrahigh
quality whispering gallery mode (WGM) resonators where counter propagating modes are coupled by scattering centers\cite{Armani03} as shown in Fig.~\ref{fig1}(b). Silica microtoroidal resonators\cite{Armani03}, and polished crystal microdisks \cite{06Grudinin33}, for example respectively exhibit high $\chi^{(3)}$ and $\chi^{(2)}$ nonlinearity, as well as strong scattering. These resonators are capable of providing strong optical confinement in a scalable, robust, microfabricated architecture with high coupling efficiency to optical fiber\cite{Armani03}. They are therefore an ideal candidate to produce, in-fiber, the strong  entanglement predicted here. Such entanglement would be an important resource for scalable quantum information networks, as well as long-distance fundamental tests of quantum mechanics such as the EPR paradox\cite{BowenRMP} and Bell tests\cite{Aspect}. Furthermore, recent research has demonstrated an exquisite capability to engineer microresonator mechanical properties\cite{kippenberg08}, providing the potential to suppress noise due to  guided wave Brillouin and Raman scattering which presents a strong constraint on existing in-fiber entanglement sources\cite{elser06}.

{\it Analysis:} In our model, within each coupled OPO the nonlinear process converts $k$ pump photons at frequency $\omega_p$ into a pair of signal and idler photons at frequencies $\omega_{1} \! = \! \Omega \! - \! \Delta^{\prime}$ and $\omega_{2} \! = \! \Omega \! + \! \Delta^{\prime}$; where to maintain energy conservation $\Omega \! = \! \frac{k}{2}\omega_p$. Since the pump fields in such systems are typically bright we treat them classically here, with $\alpha_{x}$ and $\alpha_{y}$ denoting their coherent amplitudes, and the subscripts $x$ and $y$ used throughout to distinguish the two cavity modes. Assuming the nonlinear interaction strength $\Gamma$ is identical for modes $x$ and $y$, and the coherent coupling rate $g$ is independent of detuning $\Delta'$, as is the experimentally relevant case, and taking the rotating wave approximation yields the system Hamiltonian

\begin{eqnarray}
    H^{(k+1)} \! \! \! &=& \! \hbar \displaystyle \sum_{n=1}^{2} \! \left [ g (\hat{a}_{xn} \hat{a}^{\dagger}_{yn} \! + \! \hat{a}^{\dagger}_{xn}\hat{a}_{yn}) +  \! \displaystyle \sum_m \! \omega_{mn}\hat{a}^{\dagger}_{mn} \hat{a}_{mn} \right ] \nonumber \\
    &+& \! i\hbar\Gamma\displaystyle\sum_m(\alpha_{mp}^k \hat{a}^{\dagger}_{m1} \hat{a}^{\dagger}_{m2} -
    \alpha^{k\dagger}_{mp} \hat{a}_{m1} \hat{a}_{m2}), \label{H}
\end{eqnarray}

which combines the Hamiltonians describing OPO\cite{Walls95,Grosse08}
 and coherent coupling\cite{Bowen07} used in previous work; where $m \! \in \! \{ x,y \}$, $\hat{a}$ is the annihilation operator, and the subscripts 1 and 2 denote the signal and idler fields respectively. Assuming both cavity modes $x$ and $y$ have identical input coupling and loss rates, denoted by $\gamma_{\rm in}$ and $\gamma_{l}$ respectively, as is typical for experimental systems, and
applying the quantum Langevin equation\cite{Gardiner00} to Eq.~(\ref{H}) we obtain four equations of motion for the signal and idler fields

\begin{eqnarray}
\!\dot{\hat{a}}_{x1}\!\! \! \!&=\!\Gamma
{\alpha}_{x}^k \hat{a}^{\dagger}_{x2} \!-\!(\!\gamma\!\!+\!\!i\Delta^{\!\prime})\hat{a}_{x1}\!-\!i{g}^{\prime}\!\hat{a}_{y1}\!-\! \sqrt{\! 2\gamma_{\rm{in}}}\hat{a}_{\rm{in},{x}1}\!+\!\sqrt{ \! 2\gamma_{{l}}}\hat{a}_{{ l},{x}{1}} \nonumber\\
\!\dot{\hat{a}}_{x2}\!\! \!&=\!\Gamma
{\alpha}_{x}^k \hat{a}^{\dagger}_{x1} \!-\!(\!\gamma\!\!-\!\!i\Delta^{\!\prime})\hat{a}_{x2}\!-\!i{g}^{\prime}\!\hat{a}_{y2}\!-\! \sqrt{ \! 2\gamma_{\rm{in}}}\hat{a}_{\rm{in},{ x}2}\!+\!\sqrt{\! 2\gamma_{{l}}}\hat{a}_{{l},x{2}} \nonumber\\
\!\dot{\hat{a}}_{y1}\!\! \!&=\!\Gamma
{\alpha}_{y}^k \hat{a}^{\dagger}_{y2} \!-\!(\!\gamma\!\!+\!\!i\Delta^{\!\prime})\hat{a}_{y1}\!-\!i{g}^{\prime}\!\hat{a}_{x1}\!-\! \sqrt{\! 2\gamma_{\rm{in}}}\hat{a}_{\rm{in},{ y}1}\!+\!\sqrt{ \! 2\gamma_{{l}}}\hat{a}_{{l},y{1}} \nonumber\\
\!\dot{\hat{a}}_{y2}\!\! \!&=\!\Gamma
{\alpha}_{y}^k \hat{a}^{\dagger}_{y1} \!-\!(\!\gamma\!\!-\!\!i\Delta^{\!\prime})\hat{a}_{y2}\!-\!i{g}^{\prime}\!\hat{a}_{x2}\!-\! \sqrt{\! 2\gamma_{\rm{in}}}\hat{a}_{\rm{in},{ y}2}\!+\!\sqrt{\! 2\gamma_{{l}}}\hat{a}_{{l,y{2}}} \nonumber
\end{eqnarray}

where $\hat{a}_{\rm{in}}$ and $\hat{a}_{l}$ are the fields entering through the input coupler and loss mechanisms respectively, and $\gamma \! = \! \gamma_{\rm{in}} \! + \! \gamma_{l}$ is the total decay rate of each cavity. Assuming $\alpha_x$ and $\alpha_y$ are independent of time these equations can be solved in the frequency domain by taking the Fourier transform.

The dimensionless quantities $R_xe^{ik\phi_x} \! = \! \frac{\Gamma|\alpha_{x}|^{k}}{\gamma}e^{ik\phi_x}$,
$R_ye^{ik\phi_y} \! = \! \frac{\Gamma|\alpha_{y}|^{k}}{\gamma}e^{ik\phi_y}$,
$g \! = \! g^{\prime}/\gamma$, $\Delta^{\prime} \! = \! \Delta/\gamma$, and  $\eta \! = \! \gamma_{\rm in}/\gamma$ are used; where $\eta$ is the cavity escape efficiency and $R$ is a dimensionless pump strength with the physical significance that $R=1$ corresponds to the OPO threshold.
The solution can then be expressed compactly as $\frac{1}{2}\textbf{\emph{M}}\hat{\textbf{\emph{A}}} \! = \! \sqrt{\eta} \hat{\textbf{\emph{A}}}_{\rm{in}} \! + \! \sqrt{1-\eta}\hat{\textbf{\emph{A}}}_{l}$, %

where

\begin{equation}
\textbf{\emph{M}}=\left[
\begin{array}{cccc}
i\Delta -1 & R_x e^{i k \phi_x} & -i{g} & 0 \\
R_x e^{i k \phi_x} & i\Delta-1 &0 & i{g} \\
-i{g} & 0 & i\Delta-1 & R_y e^{i k \phi_y} \\
0 & i{g} & R_y e^{-i k \phi_y} &i\Delta-1
\end{array}
\right],
\end{equation}

and throughout vectors $\hat{\textbf{\emph{A}}}_{{q}}$ are of the form  $\hat{\textbf{\emph{A}}}_{{q}} \! = \! [\hat{a}_{{q},{x}1},\hat{a}^{\dagger}_{{q},{x}2},\hat{a}_{{q},{y}1},\hat{a}^{\dagger}_{{q},{y}2}]^{\rm{T}}$.

One can then use the input/output relations\cite{Gardiner00} $\hat{\textbf{\emph{A}}}_{\rm{out}} \! = \! \sqrt{2\gamma_{\rm{in}}}\hat{\textbf{\emph{A}}} \! + \! \hat{\textbf{\emph{A}}}_{\rm{in}}$
to solve for the output fields,

\begin{equation}
    \hat{\textbf{\emph{A}}}_{\rm{out}}= [\textbf{\emph{I}}_4 + 2\eta \textbf{\emph{M}}^{-1}]\hat{\textbf{\emph{A}}}_{\rm{in}} + 2\sqrt{\eta(1-\eta)}\textbf{\emph{M}}^{-1}\hat{\textbf{\emph{A}}}_l,
    \label{main}
\end{equation}
where $\textbf{\emph{I}}_4$ is the $4 \! \! \times \! \! 4$ identity matrix.
General output field quadratures for each of the four system modes are then readily obtained as $\hat X^{\theta}(\Delta) \! = \!  \hat a(\Delta)e^{-i\theta} \! + \! \hat a^\dagger(\Delta)e^{i\theta}$.

Gaussian continuous variable entanglement between  orthogonal quadratures pairs of two light beams can be completely characterized  by the correlation matrix of the system\cite{Walls95}; which can be constructed in our case from  Eq.~(\ref{main}) for any two pairs of output fields. Here we follow the procedure of Duan {\it et. al.}\cite{Duan00} for bipartite continuous variable systems, who showed that the correlation matrix of any Gaussian state can be transformed reversibly into a standard form, from which the {\it degree of inseparability} $\mathcal{I}$ can be defined such that $\mathcal{I} \! < \! 1$ is a  necessary and sufficient condition for entanglement, with maximal entanglement characterized by $\mathcal{I} \! = \! 0$. Here we follow the method of Grosse\cite{Grosse08}, to numerically calculate the degree of inseparability from the output field correlation matrix of the coupled cavity system.

{\it Results:} In general, entanglement can be observed between the fields $\hat{a}_{{x}{1}}$, $\hat{a}_{{x}{2}}$, $\hat{a}_{{y}{1}}$, and $\hat{a}_{{y}{2}}$ through direct individual measurements on each. However, here we consider the more experimentally realistic situation of heterodyne detection performed separately on the outputs of each of the subsystems $x$ and $y$ with a local oscillator at frequency $\Omega$.

In this case for each subsystem both signal and idler fields $\hat{a}_{{m} {1}}$ and $\hat{a}_{{m} {2}}$ beat with the local oscillator and, combined, produce a signal directly proportional to the joint-quadrature operator
\begin{equation}
    \begin{array}{ccl}
    \tilde{X}_{m,t}^{\theta}(\Delta) = \frac{1}{\sqrt{2}}\left [ \hat{X}_{m,1}^{\Delta \cdot t+\theta}(\Delta)+\hat{X}_{m,2}^{-\Delta \cdot t+\theta}(\Delta) \right ],\\
    \end{array}
    \label{detection}
\end{equation}
where $m \! \in \!  \{ x,y \}$, $\theta$ is the local oscillator phase, and $t$ is the time at which the measurement occurs. As a result of the frequency difference between the signal and idler fields we see that a time delay has the effect of rotating the observed signal and idler quadratures in opposite directions.
Entanglement of the orthogonal quadrature pair $\tilde{X}^{\theta_{x}}_{{x},\tau_{x}}(\Delta)$ and $\tilde{X}^{\theta_{x}+\pi/2}_{{x},\tau_{x}}(\Delta)$ of field $x$ with the equivalent quadratures of field $y$ is possible due to the Heisenberg uncertainty relation $\Delta^{2}X^{\theta}_{m}\Delta^{2}X^{\theta+\pi/2}_{m} \! \geq \! 1$ which originates from the commutation relation $[\tilde{X}^\theta_{m},\tilde{X}^{\theta+\pi/2}_{m}] \! = \! 2i$. We analyze entanglement of this kind henceforth, and restrict our analysis to the
optimal situation where $\theta_x \! = \! \theta_y \! = \! \theta$.

\begin{figure}[t!]
\begin{center}
\includegraphics[width=7cm]{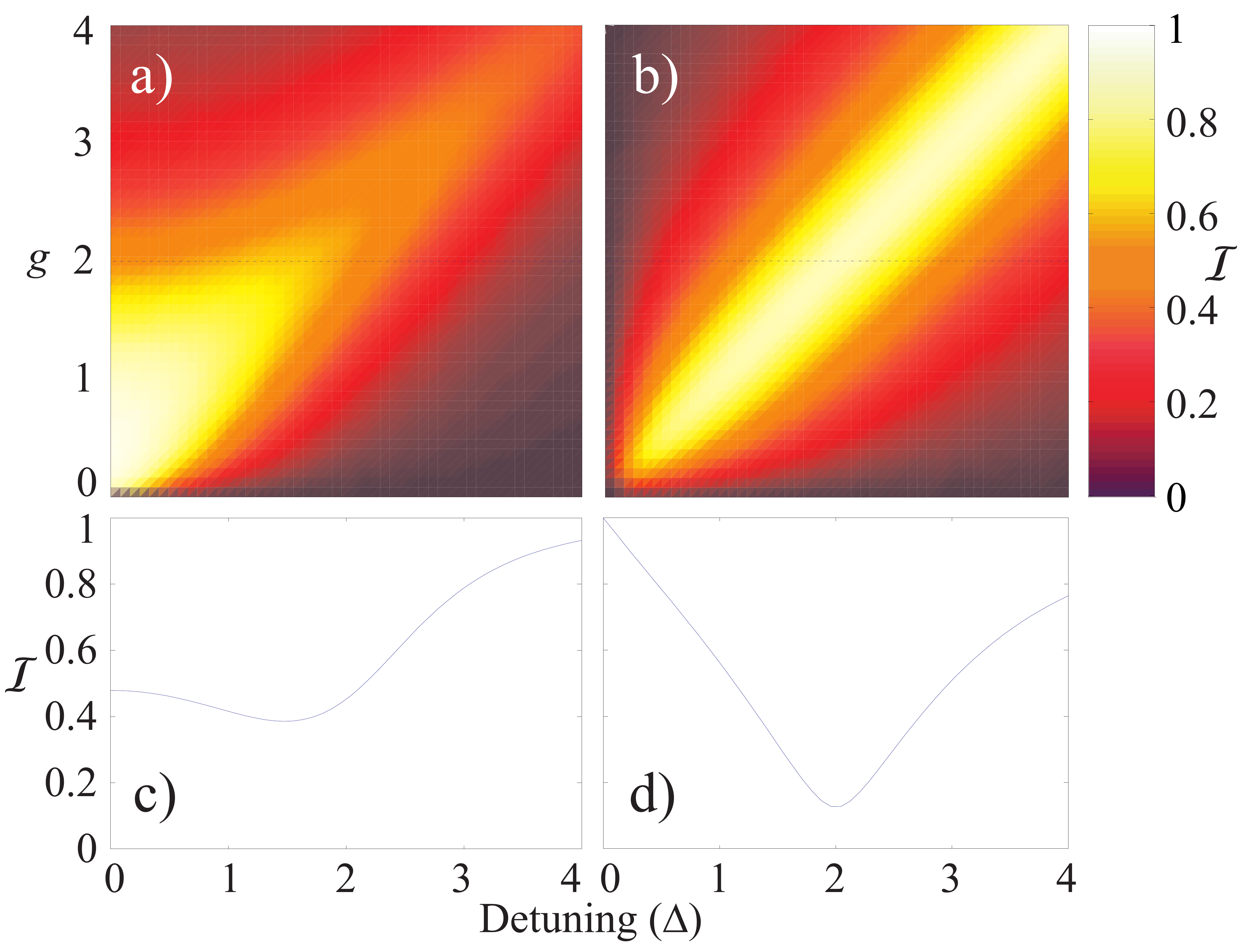}
\caption{(color online). Inseparability as a function of detuning $\Delta$ and coupling rate $g$. a) and c) synchronous entanglement with $\Delta \phi = 0$; b) and d) asynchronous entanglement with $\Delta \phi = \pi/k$. Model parameters: $R_x=R_y=0.9$ and $\eta=0.99$.}
\label{fig2}
\end{center}
\vspace{-8mm}
\end{figure}

In all nonlinear optics based entanglement sources to date, due to the near instantaneous response of the nonlinearity the quantum correlations between output fields are observed to be synchronous. Hence, here we begin by considering the case where the output fields are measured at the same time, i.e. $t_x \! = \! t_y$. Fig.~\ref{fig2}(a,c) illustrates the degree of inseparability calculated for this case as a function of coupling rate and detuning from cavity resonance, where we have chosen equal intensity in-phase pump fields and parameters typically achieved in experiments.

Strong entanglement is observed.
When $g \! \ll \! 1$ the best entanglement is near resonance as one might perhaps expect. However for large coupling one observes both that the entanglement is quenched and that the maximum entanglement shifts to nonzero detuning. The shift can be understood from the well known mode splitting that occurs in coupled cavity systems\cite{Armani03}.
As a result fields produced at frequencies detuned by $g$ from the unperturbed resonance frequencies are on resonance within the cavity and the nonlinear response is enhanced. The entanglement quenching, however, is surprising since the coupling is a reversible process. To investigate this further we consider the asynchronous
case where the measurement on field $x$ is delayed by a time $\tau \! = \! t_x \! - \! t_y$.

\begin{figure}[t!]
\begin{center}
\includegraphics[width=\columnwidth]{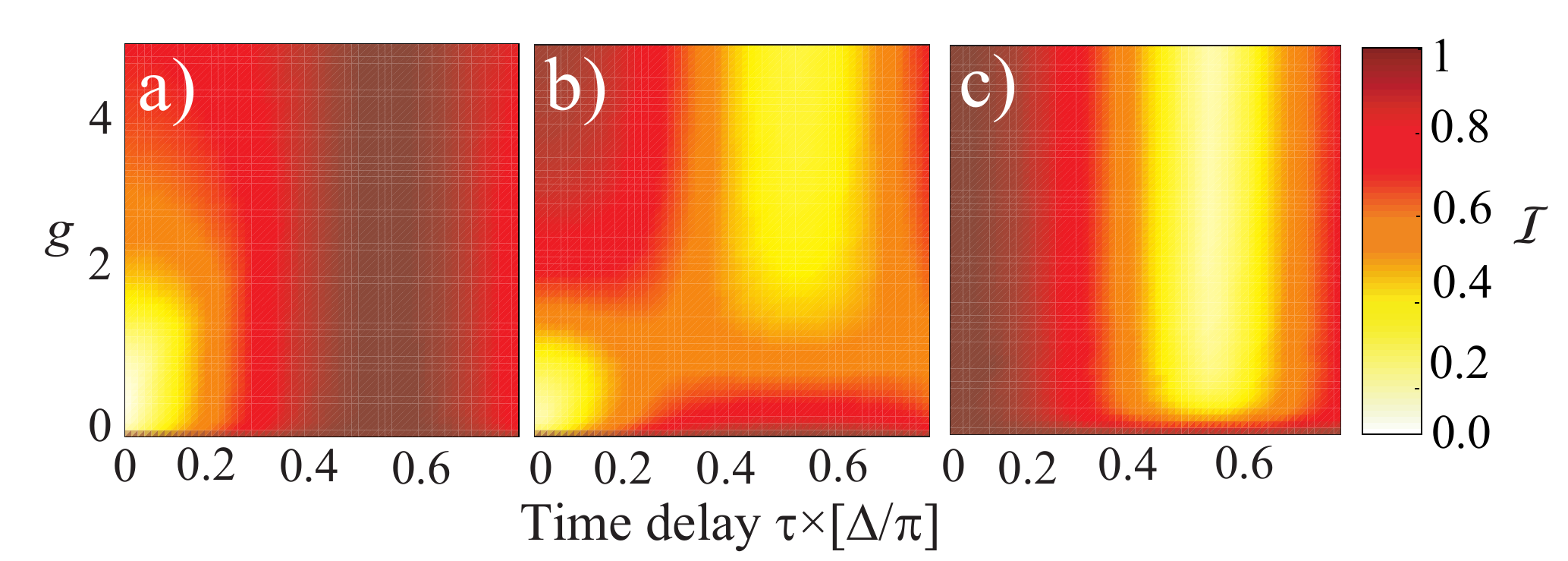}
\caption{(color online). Inseparability as a function of measurement time delay optimized over detuning. a) $\Delta \phi \! = \! 0$, b) $\Delta \phi \!  = \! \pi/2k$, c) $\Delta \phi \! = \! \pi/k$. Model parameters: $R_x \! = \! R_y \! = \! 0.9$, $\eta \!  = \! 0.99$, $g \! = \! 1$.}
\label{FigTimeDelay}
\end{center}
\vspace{-8mm}
\end{figure}

Fig.~\ref{FigTimeDelay} shows the evolution of entanglement with measurement time delay for three different relative pump phases $\Delta\phi$, where we have optimized the inseparability over detuning
$\Delta$ as is typically performed in experiments using spectral analysis. In Fig.~\ref{FigTimeDelay}(a) the pumps are in phase and we see the normal behavior for a nonlinear optics based entanglement source, with maximal entanglement in the synchronous case where $\tau \! = \! 0$; and as seen before in Fig.~\ref{fig2}(a,c) a strong dependence on $g$, with entanglement optimized at $g \! \approx \! 0.5$. In Fig.~\ref{FigTimeDelay}(b) a pump phase delay of $\pi/2k$ is introduced and we see an overall reduction in the level of entanglement. However, now as the coupling rate increases and the synchronous entanglement is quenched, a transition occurs to entanglement in an asynchronous regime with a time delay between detection events of $\tau \! = \! \pi/2\Delta$. In contrast, this asynchronous entanglement is optimized at large $g$. In Fig.~\ref{FigTimeDelay}(c) where $\Delta \phi \! = \! \pi/k$ the synchronous entanglement is no longer apparent for any  $g$, with asynchronous entanglement present for all but very small $g$. The dependence of the degree of asynchronous inseparability on coupling
 rate and detuning for this case are shown in Fig.~\ref{fig2}(b,d). It is clear that now the maximum entanglement always occurs at $\Delta \! = \! g$, can be shifted far from the resonance frequency of the cavity, and is insensitive to $g$ for $g \! \gg \! 1$. This resonance
frequency shift is technically significant, allowing the
optimal entanglement to be moved away from low
frequency noise sources such as the laser relaxation oscillation
which typically limit both entanglement strength and purity.

\begin{figure}[b!]
\begin{center}
\includegraphics[width=\columnwidth]{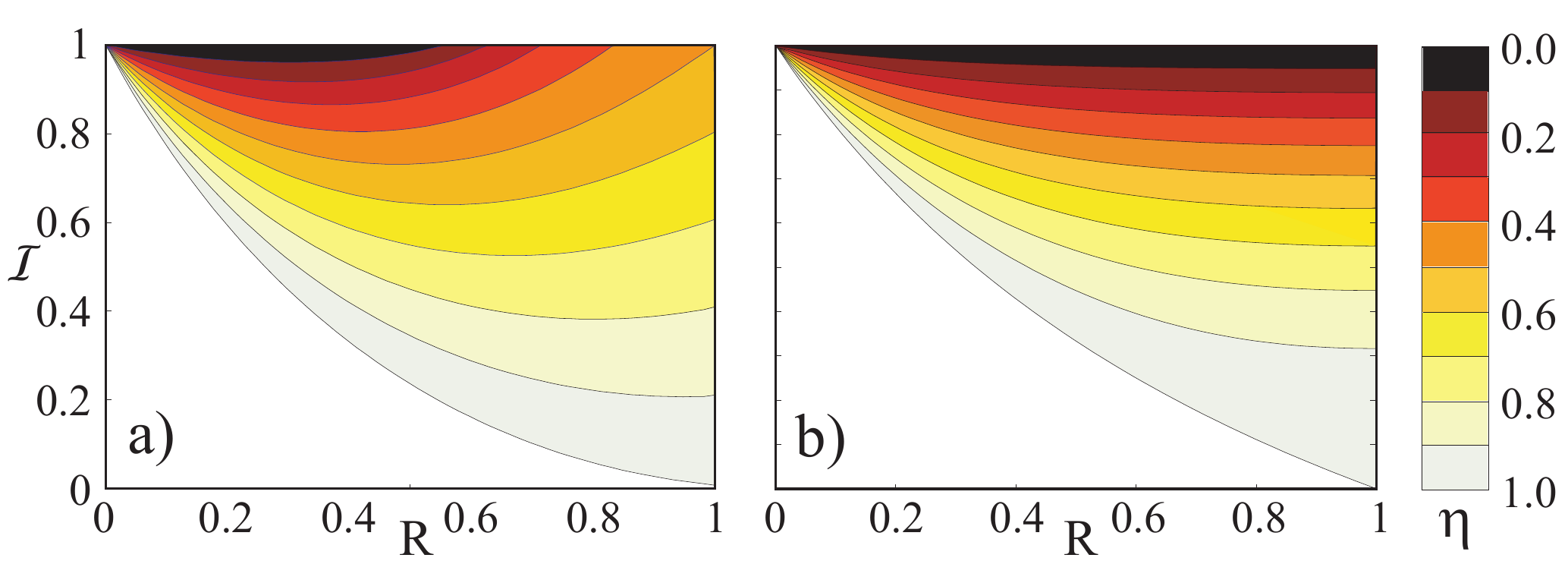}
\caption{(color online). Inseparability optimized over detuning as a function of pump strength $R=R_x=R_y$ and cavity escape efficiency $\eta$. a) synchronous entanglement with $g=0.5$ and $\Delta \phi=0$, b) asynchronous entanglement with $g \gg 1$ and $\Delta \phi = \pi/k$.}
\label{fig3}
\end{center}
\vspace{-12mm}
\end{figure}

Fig.~\ref{fig3} shows the dependence of the degree of inseparability
 on pump strength and cavity escape efficiency at optimal $\Delta$. For both synchronous and asynchronous cases,
we choose the coupling rate to provide maximal entanglement, and see
that complete entanglement is
achieved as the pump strength and escape efficiency approach unity,
although with different
functional dependance. This functional dependence will be studied
further in future work.

\begin{figure*}[tb!]
\begin{center}
\includegraphics[width=7.00in]{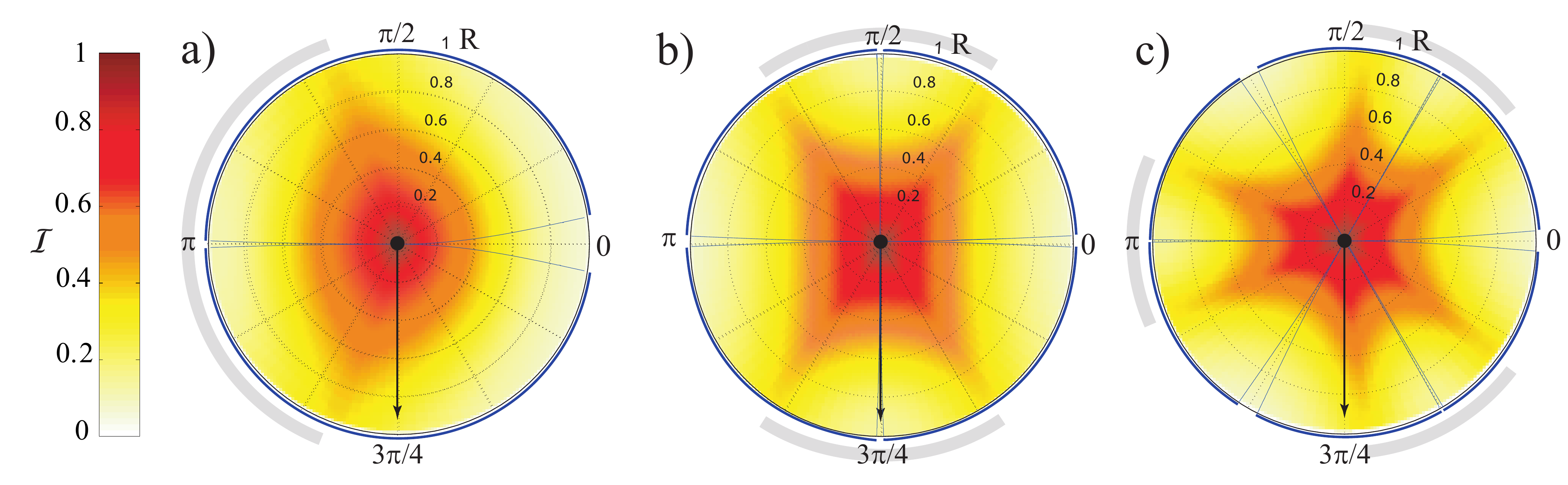}

\caption{(color online). Dependence of inseparability on pump strength $R=R_x=R_y$ (radial axis)
 and relative phase $\Delta \phi$ (azimuthal axis), optimized over
$\Delta$ and $\tau$. Order of nonlinearity: a) $\chi^{(2)}$ ($k \! = \! 1$), b) $\chi^{(3)}$
($k\! = \!2$), c) $\chi^{(4)}$ ($k\! = \!3$). Dark lines circumscribe
sectors of 4-partite entanglement; grey bands circumscribe sectors
dominated by asynchronous entanglement. Arrow: path followed for
single sided pumping. Model parameters: $\eta \! = \! 0.99$ and $g \! = \! 1$. The
plot sectors are effectively compressed by $\pi/k$ as
the order of the nonlinearity increases.}
\label{fig4}
\end{center}
\vspace{-8mm}
\end{figure*}
A standard method of generating optical entanglement is to combine a pair of squeezed fields on a 50/50 beam splitter\cite{bowen03}, with the relative phase being critical to the strength of the entanglement. In an analogous manner the coherent coupling in our scheme couples the non-classical fields generated in the two cavity modes. It is therefore to be expected that the relative pump phase $\Delta \phi$ will critically affect the entanglement achievable in this system.
Fig.~\ref{fig4} shows the minimum degree of inseparability between the output fields optimized over time delay and detuning as a function of relative pump phase and pump strength for different orders of nonlinearity. A coupling rate of $g \! = \! 1$ is chosen which, as shown in Fig.~\ref{FigTimeDelay}, is an interesting case where strong entanglement can be simultaneously generated both synchronously and asynchronously.
This represents a regime of asynchronous 4-partite entanglement, where the time-delayed and the non-time-delayed outputs of $x$ are both entangled to both time-delayed and non-time-delayed outputs of $y$. We see that 4-partite entanglement is apparent throughout the majority of the parameter space. This multipartite entanglement will be investigated further in future work.

Returning to bi-partite entanglement, it is clear from
Fig.~\ref{fig4}  that
the entanglement increases with increasing pump strength.  We also see
a strong dependence on $\Delta \phi$ as expected, however for high pump strength substantial entanglement can be generated for all $\Delta \phi$ in distinct contrast to conventional entanglement sources where typically entanglement generation fails over some range of $\Delta\phi$.

Finally we turn our attention to a robust and straightforward experimental implementation. Consider the case where only cavity mode $x$ is directly pumped with mode $y$ pumped indirectly through coherent coupling from mode $x$. In this case one finds that $\alpha_{y} \! = \! ig\alpha_{x}$. Hence the relative phase between the pumps is $\Delta\phi \! = \! 3\pi/4$, so that when $g=1$ single sided pumping tracks the downwards facing arrow in Fig.~\ref{fig4}. We see that in the special case of $\chi^{(3)}$ nonlinearity relevant to silica microtoroidal resonators\cite{Armani03} optimum entanglement can be achieved naturally in this single sided pumping scenario. For this case the simple analytical solution
\begin{equation}\small
\mathcal{I}=\sqrt{{1-\frac{16\eta R[\Delta-R(1-\eta)]}{[(1+R)^2+(\Delta-1)^2][(1-R)^2+(\Delta+1)^2]}}},
\label{sinplepump}
\end{equation}

for the inseparability can be obtained. It is clear that as $\{R,\eta,\Delta\} \! \rightarrow \! 1$, $\mathcal{I} \! \rightarrow \! 0$. Hence for sufficiently strong pumping and high efficiency, strong entanglement can be generated with only a single external pump field.

In conclusion, we have shown that strong entanglement can be generated between the output fields of a pair of coherently coupled nonlinear optical cavities.  For sufficiently strong coupling, the quantum correlations become asynchronous, and can therefore be used in conjunction with quantum teleportation to achieve an all-optical quantum memory. A straightforward experimental implementation applicable to ultrahigh quality WGM microresonators such as silica microtoroids is proposed requiring only one external pump field.

The authors thank N. Grosse for contributions to numerical
simulations. This research was funded by the Australian Research Council Discovery Project DP0987146, the MacDiarmid Institute, and New Zealand
Foundation for Research Science and Technology contract NERF-C08X0702.

\end{document}